\RequirePackage{fix-cm}

\documentclass[twocolumn]{svjour3}        
\smartqed  
\usepackage{graphicx}

\newcommand{\beq}{\begin{equation}}
\newcommand{\eeq}{\end{equation}}
\newcommand{\beqa}{\begin{eqnarray}}
\newcommand{\eeqa}{\end{eqnarray}}
\newcommand{\ba}{\begin{array}}
\newcommand{\ea}{\end{array}}

\begin{document}

\title{Reliable equation of state for composite bosons \\ 
in the 2D BCS-BEC crossover}

\author{L. Salasnich}

\institute{L. Salasnich \at
              Dipartimento di Fisica e Astronomia ``Galileo Galilei'', \\
Universita di Padova, Italy  \\
INO-CNR, Sezione di Sesto Fiorentino, Italy \\
              \email{luca.salasnich@unipd.it} }

\date{Received: date / Accepted: date}

\maketitle

\begin{abstract}
We briefly discuss recent experiments on the BCS-BEC crossover 
with ultracold alkali-metal atoms both in three-dimensional 
configurations and two-dimensional ones. Then we 
analyze the quantum-field-theory formalism used to describe 
an attractive $D$-dimensional Fermi gas taking into account 
Gaussian fluctuations. Finally, we apply this formalism 
to obtain a reliable equation of state of the 2D system at low 
temperaratures in the BEC regime of the crossover by performing a meaningful 
dimensional regularization of the divergent zero-point energy 
of collective bosonic excitations.  

\end{abstract}

\section{BCS-BEC crossover with ultracold atoms} 
 
In 2004 the {3D BCS-BEC crossover} has been 
observed with {ultracold gases made of fermionic 
$^{40}$K and $^6$Li alkali-metal atoms} \cite{exp1,exp2,exp3,exp4}. 
As schematically shown in Fig. 1, this crossover is obtained 
by changing with a Feshbach resonance 
the s-wave scattering length ${ a_F}$ of the inter-atomic potential. There 
are three characteristic regimes which depend on the value 
of the scattering length $a_F$ \cite{tedeschi}: 
\\
-- ${ a_F}\to 0^-$, that is the BCS regime of weakly-interacting 
Cooper pairs;   
\\
-- ${ a_F}\to \pm \infty$, that is unitarity limit of strongly-interacting 
Cooper pairs;  
\\
-- ${ a_F}\to 0^+$, that is the BEC regime of bosonic dimers.  

\begin{figure}[t]
\begin{center}
\includegraphics[width=.45\textwidth]{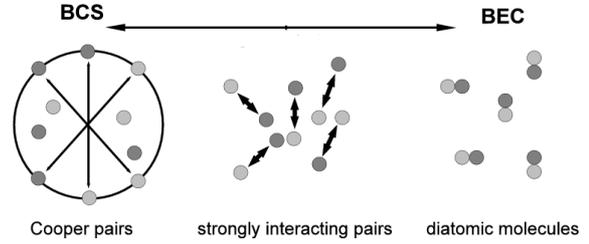}
\caption{The figure shows the evolution 
from the BCS limit with large, spatially overlapping 
Cooper pairs to the BEC limit with tightly bound molecules. 
The systems is a fermionic gas made of attractive 
two-spin-component atoms with s-wave scattering length $a_F$. 
Adapted from Ref. \cite{tedeschi}.}
\end{center}
\end{figure}

The crossover from a {BCS superfluid} (${ a_F}<0$) 
to a { BEC of molecular pairs} (${ a_F}>0$) 
has been investigated experimentally around a Feshbach resonance, 
where the s-wave scattering length $a$ diverges ($a_F=\pm \infty$), 
and it has been shown that the system is metastable \cite{exp1,exp2,exp3,exp4}. 
The detection of quantized vortices under rotation \cite{mit-vortex} 
has clarified that this dilute gas of ultracold atoms is superfluid. 
Usually the BCS-BEC crossover is analyzed in terms of 
\beq 
y = {1\over k_F { a_F}}
\eeq
the inverse scaled interaction strength, where 
$k_F=(3\pi^2n)^{1/3}$ is the Fermi wave number and 
$n$ the total fermionic density. 
The system is dilute because $r_e k_F \ll 1$, with $r_e$ the 
effective range of the inter-atomic potential.  

In 2014 also the {2D BCS-BEC crossover} has been 
achieved \cite{russi} with a {quasi-2D Fermi gas 
of $^6$Li atoms} with widely tunable s-wave interaction, 
measuring the pressure $P$ {\it vs} the gas parameter 
${ a_B} n_B^{1/2}$, with ${ a_B} = { a_F}/(2^{1/2}e^{1/4})$ 
the bosonic scattering length between molecules (see below and 
\cite{sala-flavio1}) and $n_B=n/2$ the bosonic density. 
In Fig. 2 we plot the pressure $P$
of the system as a function of the gas parameter. 

\begin{figure}[t]
\begin{center}
\includegraphics[width=.40\textwidth]{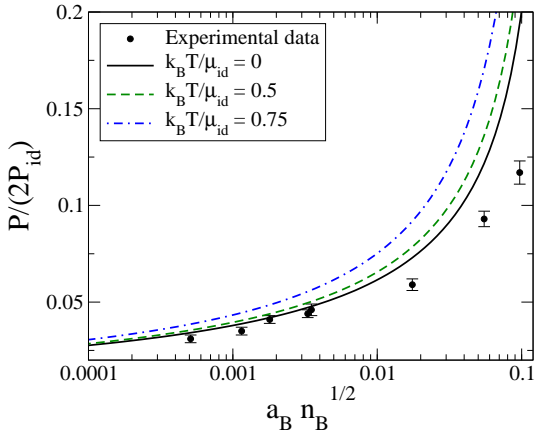}
\caption{Scaled pressure $P/(2P_{id})$ 
of the 2D Bose gas as a function of the gas parameter $a_B n_B^{1/2}$, 
where $P_{id}$ is the pressure of an 
ideal 2D gas, $a_B$ is the 
s-wave scattering length of bosons, and $n_B$ is the bosonic 2D density. 
The filled circles with error bars are the experimental data 
\cite{russi}. The curves are obtained 
with our beyond-mean-field theory (see below and \cite{sala-flavio1}).}
\end{center}
\end{figure}

Fig. 2 shows a good agreement between the experimental data \cite{russi} 
and our theoretical curves only in the deep weak-coupling 
regime $a_B n^{1/2} < 0.01$ and 
assuming a very small scaled temperature $k_BT/\mu_{id}$. 
In the next two sections we shall discuss some details 
of our beyond mean-field theory \cite{sala-flavio1,sala-flavio2}.

\section{Theory for a $D$-dimensional Fermi superfluid}

To study the attractive $D$-dimensional Fermi liquid 
we adopt the { path integral formalism} \cite{nagaosa}. 
The { partition function} ${\cal Z}$ of the uniform 
system  with fermionic fields $\psi_{s} ({\bf r},\tau )$ 
at temperature $T$, in a $D$-dimensional volume $L^D$, 
and with chemical potential $\mu$ reads
\beq 
{\cal Z} = \int {\cal D}[\psi_{s},\bar{\psi}_{s}] 
\ \exp{\left\{ -{1\over \hbar} \ S  \right\} } \; , 
\eeq
where ($\beta \equiv 1/(k_B T)$ with $k_B$  Boltzmann's constant)
\beq 
S = \int_0^{\hbar\beta} 
d\tau \int_{L^D} d^D{\bf r} \ {\cal L}
\eeq
is the { Euclidean action functional} 
with { Lagrangian density} 
\beq 
{\cal L} = \bar{\psi}_{s} \left[ \hbar \partial_{\tau} 
- \frac{\hbar^2}{2m}\nabla^2 - \mu \right] \psi_{s} 
+ {g} \, \bar{\psi}_{\uparrow} \, \bar{\psi}_{\downarrow} 
\, \psi_{\downarrow} \, \psi_{\uparrow} 
\eeq
where ${g}$ is the attractive strength (${g}<0$) 
of the s-wave coupling. 

Through the usual { Hubbard-Stratonovich transformation} 
the Lagrangian density ${\cal L}$, 
quartic in the fermionic fields, 
can be rewritten  as a quadratic form by introducing the
{ auxiliary complex scalar field} $\Delta({\bf r},\tau)$ so that:
\beq 
{\cal Z} = \int {\cal D}[\psi_{s},\bar{\psi}_{s}]\, 
{\cal D}[\Delta,\bar{\Delta}] \ 
\exp{\left\{ - {S_e(\psi_s, \bar{\psi_s},
\Delta,\bar{\Delta}) \over \hbar} \right\}} \; , 
\eeq
where 
\beq 
S_e(\psi_s, \bar{\psi_s},\Delta,\bar{\Delta}) = \int_0^{\hbar\beta} 
d\tau \int_{{L^D}} d^D{\bf r} \ 
{\cal L}_e(\psi_s, \bar{\psi_s},\Delta,\bar{\Delta})
\eeq
and the (exact) effective Euclidean Lagrangian 
density ${\cal L}_e(\psi_s, \bar{\psi_s},\Delta,\bar{\Delta})$ reads 
\beqa 
{\cal L}_e &=&
\bar{\psi}_{s} \left[  \hbar \partial_{\tau} 
- {\hbar^2\over 2m}\nabla^2 - \mu \right] \psi_{s} 
\nonumber 
\\
&+& \bar{\Delta} \, \psi_{\downarrow} \, \psi_{\uparrow} 
+ \Delta \bar{\psi}_{\uparrow} \, \bar{\psi}_{\downarrow} 
- {|\Delta|^2\over {g}} \; . 
\label{ltilde}
\eeqa

We want to investigate the effect of 
fluctuations of { the gap field} $\Delta({\bf r},t)$ around its
mean-field value $\Delta_0$ which may be taken to be real. 
For this reason we set 
\beq 
\Delta({\bf r},\tau) = \Delta_0 +\eta({\bf r},\tau) \; , 
\label{polar}
\eeq
where $\eta({\bf r},\tau)$ is the complex field which describes 
pairing fluctuations. 

In particular, we are interested in { the grand potential} $\Omega$, 
given by 
\beq 
\Omega = - {1\over \beta} \ln{\left( {\cal Z} \right)} \simeq - {1\over \beta} 
\ln{\left( {\cal Z}_{mf} {\cal Z}_g \right)} = \Omega_{mf} + \Omega_{g} 
\; , 
\eeq
where 
\beq 
{\cal Z}_{mf} = \int {\cal D}[\psi_{s},\bar{\psi}_{s}]\, 
\exp{\left\{ - {S_e(\psi_s, \bar{\psi_s}, 
\Delta_0) \over \hbar} \right\}} \;  
\eeq
is the mean-field partition function and 
\beq
{\cal Z}_g = \int {\cal D}[\psi_{s},\bar{\psi}_{s}]\, 
{\cal D}[\eta,\bar{\eta}] \ 
\exp{\left\{ - {S_g(\psi_s, \bar{\psi_s},
\eta,\bar{\eta},\Delta_0) \over \hbar} \right\}} 
\eeq
is the partition function of Gaussian pairing fluctuations. 

To make a long story short, 
one finds that in the gas of paired fermions there are 
{ two kinds of elementary excitations} \cite{nagaosa,sala-solo,sala-berto}: 
{ fermionic single-particle excitations} with energy 
\beq 
E_{sp}(k)=\sqrt{\left({\hbar^2k^2\over 2m}-\mu\right)^2+\Delta_0^2} \; ,  
\label{ex-fermionic}
\eeq 
where $\Delta_0$ is the pairing gap, and { bosonic collective 
excitations} with energy 
\beq 
E_{col}(q) = \sqrt{{\hbar^2q^2\over 2m} \left( \lambda \ {\hbar^2q^2\over 2m} 
+ 2 \ m \ c_s^2 \right)} \; , 
\label{ex-bosonic}
\eeq
where $\lambda$ is the first correction 
to the familiar low-momentum phonon dispersion 
$E_{col}(q) \simeq c_s \hbar q$ and $c_s$ is the sound velocity. Notice that 
both $\lambda$ and $c_s$ depend on the chemical potential $\mu$ 
\cite{sala-berto}.

Moreover, at the Gaussian level, the { total grand potential} reads 
\cite{nagaosa,sala-berto} 
\beq 
\Omega = \Omega_{mf} + \Omega_g \; ,  
\eeq
where 
\beq 
\Omega_{mf} = - {\Delta_0^2\over {g}}\, L^D 
+ \Omega_{F}^{(0)} + \Omega_{F}^{(T)} 
\eeq 
is the { mean-field grand potential} with 
\beq 
\Omega_{F}^{(0)} = - \sum_{\bf k} \left( E_{sp}(k) - {\hbar^2k^2\over 2 m} 
+ \mu \right) 
\eeq 
the zero-point energy of { fermionic single-particle excitations}, 
\beq 
\Omega_{F}^{(T)} = {2\over \beta } 
\sum_{\bf k} \ln{(1+e^{-\beta\, E_{sp}(k)})} \;  
\eeq
the finite-temperature grand potential of the { fermionic 
single-particle excitations}.  

The { grand-potential of Gaussian fluctuations} reads  
\beq 
\Omega_g =  \Omega_{g,B}^{(0)} + \Omega_{g,B}^{(T)} \; , 
\eeq
where 
\beq 
\Omega_{g,B}^{(0)} = {1\over 2} \sum_{\bf q} E_{col}(q) 
\eeq 
is the zero-point energy of { bosonic collective excitations} and 
\beq 
\Omega_{g,B}^{(T)} = {1\over \beta } 
\sum_{\bf q} \ln{(1- e^{-\beta\, E_{col}(q)})} \;  
\eeq
is the finite-temperature grand potential of the { bosonic collective 
excitations}. 

Both $\Omega_{F}^{(0)}$ and 
$\Omega_{g,B}^{(0)}$ are ultraviolet divergent in any dimension $D$ 
($D=1,2,3$) and the regularization of these divergent terms 
is complicated by the fact that one also must take into account 
the BCS-BEC crossover \cite{sala-berto,sala-flavio1}. 

\section{Results of the two-dimensional Fermi superfluid} 

In the analysis of the {two-dimensional attractive Fermi gas} 
one must remember that, contrary to the 3D case, 
{ 2D realistic interatomic attractive potentials have always a bound state}. 
In particular, 
the binding energy ${ \epsilon_b}>0$ of two fermions can be written 
in terms of the positive 2D fermionic scattering length ${ a_F}$ as 
\beq 
{ \epsilon_b}= {4\over e^{2\gamma}}{\hbar^2\over 
m { a_F}^2} \; ,   
\label{eb-af}
\eeq
where $\gamma=0.577...$ is the Euler-Mascheroni constant \cite{mora2003}. 
Moreover, 
the attractive (negative) interaction strength ${g}$ 
of s-wave pairing is related to the 
binding energy ${ \epsilon_b}>0$ of a fermion pair in vacuum by 
the expression \cite{randeria1989} 
\beq 
- \frac{1}{g} 
= \frac{1}{2L^2} \sum_{\bf k} \frac{1}{{\hbar^2k^2\over 2m} + 
\frac{1}{2} { \epsilon_b}} \; . 
\label{g-eb}
\eeq 

In the {2D BCS-BEC crossover}, at zero temperature ($T=0$) the mean-field grand 
potential $\Omega_{mf}$ can be written as \cite{sala-berto,randeria1989} 
\beq 
\Omega_{mf} = - {m L^2\over 2\pi \hbar^2} 
(\mu + {1\over 2} { \epsilon_b} )^2  \;  
\label{omega-mf}
\eeq
with ${ \epsilon_b}>0$. Using 
\beq
n = - {1\over L^2} {\partial \Omega_{mf}\over \partial \mu}
\eeq
one immediately finds the chemical potential $\mu$ as a 
function of the number density $n=N/L^2$, i.e. 
\beq 
\mu = {\pi \hbar^2 \over m} n - {1\over 2} { \epsilon_b} \; . 
\label{echem-mf}
\eeq
In the BCS regime, where ${ \epsilon_b} \ll \epsilon_F$ 
with $\epsilon_F=\pi\hbar^2n/m$, 
one finds $\mu \simeq \epsilon_F >0$ while in the BEC regime, 
where ${ \epsilon_b} \gg \epsilon_F$ one has 
$\mu \simeq - { \epsilon_b}/2 <0$. 

Performing { dimensional regularization} of Gaussian fluctuations, 
we have recently found \cite{sala-flavio1} 
that the zero-temperature total grand potential is 
\beq 
\Omega = \Omega_{mf} + \Omega_{g} = - {m L^2\over \pi\hbar^2}  
(\mu + {1\over 2}{ \epsilon_b})^2 \ 
\ln{\left({{ \epsilon_b}\over 
2 (\mu + {1\over 2}{ \epsilon_b}) } \right)} \; .  
\eeq 
in the deep BEC regime. 
Introducing $\mu_B = 2(\mu + \epsilon_b/2)$ as the chemical potential 
of composite bosons with mass $m_B=2m$ and 
density $n_B=n/2$, the zero-temperature 
total grand potential can be rewritten as 
\beq 
\Omega = - {m_B L^2\over 8\pi\hbar^2} 
\mu_B^2 \ \ln{\left({{ \epsilon_0}\over \mu_B } \right)} \; ,  
\eeq
that is exactly the Popov equation of state of 2D 
weakly-interacting bosons \cite{popov} 
provided that we identify the parameter
\beq 
{ \epsilon_0} =  {4\over e^{2\gamma+1/2}}{\hbar^2\over m_B { a_B}^2}
\eeq
of the Popov theory of bosons with  
scattering length ${ a_B}$ \cite{mora2009} 
with the binding energy 
\beq 
{ \epsilon_b} = {4\over e^{2\gamma}}{\hbar^2\over m { a_F}^2} 
\eeq
of paired fermions with scattering length 
${ a_F}$ \cite{mora2003}. Thus, we find \cite{sala-flavio1}
\beq 
\framebox[4cm]{
${ a_B} = {1\over 2^{1/2}e^{1/4}} \ { a_F} \; .$
}
\eeq
The value ${ a_B}/{ a_F}= 1/(2^{1/2}e^{1/4}) \simeq 0.551$ 
is in full agreement with other theoretical predictions: 
${ a_B}/{ a_F}=0.56$ obtained from four-body scattering 
theory \cite{petrov}, ${ a_B}/{ a_F}=0.55(4)$ obtained by 
Monte Carlo calculations \cite{bertaina}, 
and ${ a_B}/{ a_F}=0.56$ obtained very recently 
by using Gaussian fluctuations with convergence-factor 
regularization \cite{cinesi}. 

At finite temperature ($T\neq 0$) the pressure $P$ is immediately 
obtained using the thermodynamic formula $P = - \Omega/L^2$. 
Taking into account that the main thermal contribution 
is due to collective bosonic excitatons, we obtain  
\cite{sala-flavio2} from Eqs. (20) and (27) the finite-temperature 
pressure  
\beq 
P =  {m_B \over 8\pi\hbar^2} \, \mu_B^2 
\left[ \ln{\left({{ \epsilon_0}\over \mu_B } \right)} + 4 \zeta(3) 
\left( k_B T\over \mu_B \right)^3 \right] \; , 
\label{p-mio}
\eeq
and also, by using $n_B=\left({\partial \Omega\over 
\partial \mu_B}\right)_{T,L^2}$, the bosonic density 
\beq 
n_B =  {m_B \over 4\pi\hbar^2} \, \mu_B  
\left[ \ln{\left({{ \epsilon_0}\over \mu_B \, e^{1/2}} \right)} - 2 
\zeta(3) \left( k_B T\over \mu_B \right)^3 \right] \;  
\label{n-mio}
\eeq
where $\zeta(x)$ is the Riemman zeta fuction and $\zeta(3)=1.20205$. 
Eqs. (\ref{p-mio}) and (\ref{n-mio}) give, at fixed $k_BT/\mu_B$, 
a parametric formula for the the pressure $P$ 
as a function of the density $n_B$ where $\mu_B$ is the 
dummy parameter (see Fig. 2). Thus, we have a reliable 
equation of state for composite bosons in the 2D BEC-BEC 
crossover at low temperatures, i.e. when the system is 
well below the Berezinsky-Kosterlitz-Thouless critical 
temperature of the superfluid-normal transition \cite{nagaosa}. 

\section{Conclusions}

We have shown that 
the $D$-dimensional superfluid Fermi gas in the BCS-BEC crossover 
has a { divergent} { zero-point energy} due to 
{ fermionic single-particle excitations} (mean-field) and  
{ bosonic collective excitations} (Gaussian fluctuations). 
However, the regularization of the { divergent} { zero-point energy} 
gives remarkable analytical results for composite bosons 
in two dimensions \cite{sala-flavio1}: 
a reliable 2D equation of state and 
an analytical formula connecting the scattering length ${ a_B}$ 
between composite bosons and the scattering ${ a_F}$ between fermionic 
atoms. Finally, we notice that also in three-dimensions one can regularize 
the divergent zero-point energy due to fermionic and bosonic 
excitations \cite{pieri,hu,diener}. In particular, 
by performing a cutoff regularization and renormalization of Gaussian 
fluctuations, we have found very recently \cite{sala-mino} 
that $a_B=(2/3)a_F$ for composite bosons in the 3D BCS-BEC crossover.  

\begin{acknowledgements}

This work was partially supported by MIUR through the PRIN Project "Collective 
Quantum Phenomena: from Strongly-Correlated Systems to Quantum Simulators". 

\end{acknowledgements}

\end{document}